%macropackage= phyzzx
\input phyzzx
\overfullrule=0pt

\def\dslash{\not{\hbox{\kern-2pt $\partial$}}}
\def\Dslash{\not{\hbox{\kern-2pt $D$}}}
\def\Qslash{\vert{\hbox{\kern-5pt Q}}}
\def\Rslash{\vert{\hbox{\kern-5.5pt R}}}
\def\hslash{\not{\hbox{\kern+1.5pt h}}}

\def\({\left\lbrack}           \def\){\right\rbrack}
\def\{{\left\lbrace}           \def\}{\right\rbrace}

\def\pa{\partial}

\def \pa{\partial}

 \def\pa{\partial}
\def\bpa{\bar\partial}

\def\Q{$QCD_2$\ }

\def\snc{$SU(N_c)$}

%%%%%%%%%%%%%%%

%%%%%%%%%%%%%%%

\def\pl#1{{\it Phys. Lett.} {\bf #1B}}

\def\prd#1{{\it Phys. Rev.} {\bf D#1}}

\def\np#1{{\it Nucl. Phys.} {\bf B#1}}

\def\jmath#1{{\it J. Math. Phys.} {\bf #1}}

%********************
\def\nc{N_c}
\def\nf{N_f}
\def \j{h^{-1}\pa h }
\def \bj{ h\bpa h^{-1}}
%*************
\REF\DRS{S. Dimopolus, S. Rabi and L. Susskind \np {173} (1980) 208.}
\REF\GKMS{D. J. Gross, I. R. Klebanov, A. V. Matysin and A. Smilga,
``Screening vs. Confinement in 1+1 dimensions",
\np {461} (1996) 109.}
\REF\FS {Y. Frishman and J. Sonnenschein, \np  {294} (1987) 801 ;
{\it Physics Reports} {\bf 223 \# 6}  (1993)
309.}
\REF\EFHK{J.~Ellis, Y.~Frishman, A.~Hanany and M.~Karliner,
``Quark solitons as constituents of hadrons,''
hep-ph/9204212, Nucl.\ Phys.\ B{\bf 382} (1992) 189.}
\REF\Abda{W. Krauth and M. Staudacher, \pl {388} (1996) 808;
E. Abdalla, R. Mohayaee
``Quasi-integrability and two-dimensional QCD"
hep-th/9610059}
\REF\FHS{ Y. Frishman, A.Hanany  and J. Sonnenschein, Subtleties in
QCD in Two Dimensions, {\it Nucl. Phys.}\  {\bf B429}  (1994) 75.}
\REF\DFS{G. D. Date, Y. Frishman, and J. Sonnenschein,
{\it Nucl. Phys.}\  {\bf B283}  (1987) 365.}
\REF\tHooft{G. `t Hooft \np {75} (1974) 461.}
\REF\Adj{D. Kutasov, \np {414} (1994) 33;
G. Bhanot, K.Demeterfi and I. Klebanov, \prd{48} (1993) 4980-4990.}
\REF\usss  {O. Aharony, O. Ganor, J. Sonnenschein and S.
Yankielowicz, \np {399} (1993) 560.}
\REF\KuSc{D. Kutasov and A. Schwimmer, \np {442} (1995) 447.}
\REF\Adi{A.Armoni, J. Sonnenschein``Screening in the large $N_f$
$QCD_2$ and $SYM_2$" in preparation}

%%%%%%%%%%%%%%%%%%%%%%%%%%%%%%%%%%%%
\titlepage
\baselineskip=20pt
\rightline {WIS-47/96-Dec}
\rightline{TAUP-2394-96 }

\title{$QCD_2$-Screening, Confinement and Novel non-abelian Solutions}
\author {Y. Frishman \footnote{*}
{Supported in part by the Israel Academy of Sciences.}}
\address{Department of Particle Physics \break
Weizmann Institute of Science \break
 76100 Rehovot Israel}
\author { J.~Sonnenschein
\footnote{\dagger}{Work supported in part
by the ``Einstein center for theoretical physics" at the Weizmann
Institute,
the  US-Israel Binational Science Foundation and the Israel
Science foundation.}}
\address{ School of Physics and
Astronomy\break Beverly and Raymond Sackler \break
Faculty    of Exact Sciences\break
Ramat Aviv Tel-Aviv, 69987, Israel}
\abstract{We analyze the question of screening versus confinement
in bosonized massless
QCD in two dimensions. We deduce the screening behavior
of massless $SU(N_c)$ QCD with flavored fundamental fermions and fermions
in the adjoint representation.
 This is done by computing the potential between external
quarks as well as by bosonizing also the external sources and
 analyzing the states of the combined system.
We  write down novel ``non-abelian Schwinger
like" solutions of the equations of motion, compute their masses and
argue that an exchange of massive modes of this type 
 is associated with  the screening mechanism.}

\section{Introduction}
The question of confinement versus screening in four dimensional (4D)
non-abelian gauge thoeries
is one of the major problems of high-energy theory. Two dimensional (2D) gauge
theories may serve as a laboratory in exploring that problem. Just as in 4D,
also in 2D one can use the potential between two heavy
external charges, the expectation value of a Wilson loop
and the structure of the bound state spectrum  as probes of
confinement. It is believed that a potenial growing linearly at large
separation distances, an area law behaviour of the Wilson loop
and a spectrum  which is independent of the number of colors $N_c$
indicate that the system is in a confining state.
Note that in some cases, like with Higgs in the fundamental, screening and
confinement are in one phase by ``complementarity"\refmark\DRS

In a  paper by D.Gross et al.\refmark\GKMS\  it was argued that
there is a screening effect between  heavy external charges
induced by massless  dynamical fermions even if the latter are in a
representation which  has zero ``$N_c$-ality", namely, vanishing center
($Z_{N_c}\sim 0$).
%larger than that of the probes.
It was further
shown that confinement is restored  as soon as the dynamical fermions get some
non-trivial mass. In that paper both the nature of the potential and the
Wilson loop were determined in the abelian theory and in several non-abelian
cases. In one case the group was $SU(N_c=2)$ with the dynamical fermions
in {\bf 3}, and the other case was $SU(3)$ with {\bf 8} fermions. In the
latter it was shown in fact that the spinor {\bf 8} of $SO(8)$ are screening.
%\Tcl

 The potential between the external quarks can be extracted using
several different methods:
 (i) Deriving  the effective Lagrangian
(integrating out the fermions)  and then extracting
 the potential   using the static gauge
configurations that solve   the  corresponding equations of motion.\refmark\GKMS
(ii) Using  the gauge configuration that solves  the equations of motion
of the  bosonized action\refmark\FS.
(iii) Eliminating the gauge fields
from the bosonized gauged action,   solving  for static
solutions of the currents and deducing the potential as the difference between
the Hamiltionian of the systems with and without the external sources.

In the present paper we use  method
(ii)  to prove the statements of screening for massless
dynamical fermions.

Another approach to determine if the system confines is based on
  bosonizing   also the heavy external
charges.\refmark{\EFHK}
Confinement manifests itself in this double bosonized model
by the absence of soliton solutions that correspond to unbounded
quarks. In case that there are  quark finite-energy static solutions,
one may conclude that the system is non-confining. We use
also this type of
analysis for both abelian and non-abelian gauge theories.

The screening mechanism in the massless Schwinger model could be
attributed  to the  exchange of the emerged
massive photon, which   is the
only state in the exact spectrum.  The non-abelian counterpart is
clearly  much more complicated and  seems to be
a non-integrable model\refmark\Abda. However, by
introducing flavor degrees of freedom one can  pass, in the limit of
large number of flavours $N_f$ (with finite $N_c$),
to a domain where the non abelian theory resembes a collection of
$N_c^2-1$ abelian theories. In that limit the spectrum includes
$N_c^2-1$ massive modes of the type that exist in the  Schwinger model .
One can then draw  an intuitive picture of screening due to those  modes
in a similar manner to the one in the abelian theory.
As a matter of fact, it is only in the large $N_f$ limit, that one can
justify  relating  solutions of  the equations of motion
and physical states and deducing conclusions about the sctructure
of the spectrum.
One might find in  the
``massive gauge states"  an indication of the
``non-confining" structure of the spectrum. The reason for that is,
that even though they are gauge invariant states,  they
are in the adjoint representation of a ``global
color symmetry" and not singlets of that group.
These states
  had  already  been
pointed  out in an earlier work,\refmark\FHS
based on a BRST analysis and a special parametrization of the
gauge configurations. However, in that paper  we were not able
to rigorusly show that  indeed they were part of the BRST cohomology.
Note, that even if they are not in space of physical states, the
massive states could nevetheless be responsible for the
screening potential.

When passing from a screening picture at large $N_f$ and finite $N_c$
to the domain of a small number of flavors, one can anticipate
two types of  scenarios: (i) A smooth transition
 where  the screening behavior persists all the way down to $N_f=1$; (ii)
A phase transition at a certain value of $N_f$ and a confining nature below it.
One may argue that the massive modes of large $N_f$ are an artifact
of the abelianization of the theory.
To check that possibility we have searched for non-abelian solutions
of the equations of motions. Indeed, we found new non-abelian solutions
that are also  massive and are associated with the gauge fields,
namely, are in the adjoint of the global color group. We thus conclude
that this nature does not stem from the abelianization of the large
$N_f$ limit, and hence may
 present certain evidence in favour of  option (i).
A different ``patch" of the space of physical states for finite
$N_c$ and finite $N_f$, that of the low lying baryonic states, was
determined in the semiclassical domain in [\DFS]. Since those baryons
are all color singlets this might seem as a contradition to the
screening  nature of the spectrum. In fact there is no contradiction
since the baryons were discovered only for massive quarks and not for
massless ones. As was shown in [\GKMS] turning on a mass term for
the  quarks changes the picture dramatically into a confinig one.

%We present exact solutions of the equations of motion that admit a non-abelian
%behaviour. Those solutions are shown to be massive with a multipicity of
%$N_c^2-1$.
't Hooft solved the spectrum of $QCD_2$ in the large $N_c$ limit.\refmark\tHooft
In that analysis the  quarks were flavorless and
in  the color fundamental representation.
This procedure was recently also applied for adjoint fermions.
Both in the original work as well as in those of ref. [\Adj]
there is no trace of
the  massive  modes that our work analyzes. Differently stated the large
$N_c$ approach reveals a confinig spectrum both for the case of massless and
massive quarks. We believe that this is an artifact of the  large $N_c$
limit and at finite $N_c$ the spectrum of a theory with/without quark
mass behaves like a confining/screening spectrum respectively.

The paper is organized in the following way. In section 2 we review the rules
of  bosonization of two dimensional QCD with both massive and massless
fermions which transform in the fundamental or  adjoint representations.
The equations of motions of bosonized $QCD_2$ in the presence of external
currents  are derived and discussed in section 3. Non-abelian solutions
of the equations
as well as some interesting abelian ones are presented in section 4
for  the model without external sources, and in section 5 when the
latter are turned on.  The energy-momentum tensor and the spectrum that
corresponds to the solutions of sections 4 and 5 are derived in section
6. Section 7 is devoted to the analsis of the system by bosonizing also
the external currents. In section 8 we summarize  the results of the
present  work, state our understanding of the nature of  $QCD_2$ in the
different  regimes and raise  some further intersting open questions.
In appendix A we find some
 ``truly non-abelian"  $SU(N_c=2)$ solutions of  the equations
of motion of   bosonized $QCD_2$.
Appendix B is devoted to the derivation of the energy momentum tensor
of a bosonizied fermion coupled to an abelian gauge field
 and its non-abelian generalization.
In appendix C we show that  minimal energy solutions of the massive WZW model
are necessarly in a diagonal form.

\section{ Review of Bosonization in $QCD_2$ }
We start with reviewing the bosoization
formulations of $QCD_2$
with fermions in the fundamental and adjoint representations.

\subsection{  Dirac fermions in the fundamental representation.}

Multiflavor massive  \Q\ with fermions in the fundamental representation
 was
shown \refmark\DFS\ to be described
by the following action
$$\eqalign{S_{QCD_2}=&S_1(u)-{1\over
2\pi} \int d^2 z Tr(iu^{-1}\pa u \bar A + iu\bar \pa
u^{-1} A + \bar A u^{-1}A u-A \bar A )\cr
+&{m^2\over{2\pi}}\int d^2z :Tr_G[u + u^{-1}]:
+{1\over e_c^2}\int d^2 zTr_H[F^2]\cr}\eqn\mishwzw$$
where $u\in U(N_f\times N_C)$; $S_k(u)$ is a level $k$ WZW model;
$$S_k(u)={k\over 8\pi}\int d^2xTr(\partial_\mu u\partial^\mu u^{-1})
+ {k\over12\pi}\int_B d^3y\varepsilon^{ijk}Tr(u^{-1}\partial_iu)
(u^{-1}\partial_ju)(u^{-1}\partial_ku) \eqn{\mishcb}$$
$A$ and
$\bar A$ take their values in the algebra of $H\equiv SU(N_C)$; $F=\bar\pa
A-\pa\bar A+i[A,\bar A]$; $m^2$ equals $m_q\mu C$, where $\mu$ is the
normal ordering mass and $C={1\over2}e^\gamma$ with $\gamma$ Euler's
constant. Notice that the  space-time has a Minkowski signature, and  
we use a notation in which the light-cone components of
a vector $B_{\mu}$ are denoted by $B\equiv B_+$ and $\bar B\equiv B_-$.

The action for massless fermions can be simplified using the following
parametrization  $u \equiv ghle^{i\sqrt{4 \pi\over N_CN_f}\phi}$
where $h\in SU(\nc)$, $g\in SU(\nf)$ and $l\in {U(\nc\nf)\over [SU(\nc)]_{\nf}
\times
[SU(\nf)]_{\nc}\times U_B(1)}$. The action in terms of these variable takes the
form\refmark\FS
$$\eqalign{S &=S_{N_f}(h) -{\nf\over
2\pi} \int d^2 z Tr(ih^{-1}\pa h \bar A + ih\bar \pa
h^{-1} A + \bar A h^{-1}A h-A \bar A )\cr
&+S_{N_C}(g)+{1\over2\pi}\int d^2z[\pa\phi\bpa\phi]\cr
&+{1\over e_c^2}\int d^2 zTr_H[F^2]\cr}\eqn\mishwzwghl$$
Notice that the action is independent of $l$.\refmark\FS
Recall that to discuss the quark soliton structure we need the
massive $u\in U(N_f\times N_C)$ description.\refmark\EFHK

\subsection{ Majorana fermions in the adjoint representation.}

A non-abelian bosonization of Majorana fermions in the adjoint
representation\refmark\usss
can be expressed in terms of $S(h_{ad})$ where $h_{ad}$ are
$(\nc^2-1)\times (\nc^2-1)$ matrices, so that the action for the corresponding
\Q now reads, in the massive case,
$$\eqalign{S &=\half S(h_{ad}) -{1\over
4\pi} \int d^2 z Tr(ih_{ad}^{-1}\pa h_{ad} \bar A + iu\bar \pa
h_{ad}^{-1} A + \bar A h_{ad}^{-1}A h_{ad}-A \bar A )\cr
%&+S_{N_C}(g)+{1\over2\pi}\int d^2z[\pa\phi\bpa\phi]\cr
&+{m^2\over{2\pi}}\int d^2z\Tr_G:[h_{ad}
+h_{ad}^{-1} ]:\cr
&+{1\over e_c^2}\int d^2 zTr_H[F^2]\cr}\eqn\mishwzwad$$
The factor $\half$ in front of the $S(h_{ad})$ term comes from the reality
nature
of the Majorana fermions.
It is straightforward to realize that the conformal anomaly of this model is
indeed $c=\half (\nc^2-1)$, the associated  currents have affine Lie algebra
with an anomaly of $\nc$ and that the conformal dimension of $h_{ad}$ is
$\Delta h_{ad}=\half$ (left and right dimensions and a total conformal
dimension of 1).

\section{ Equations of motion  of \Q in the presence of external currents}
The equation of motions which follow from the variation of the action
\mishwzwghl\   with respect to $h$ are given
for the massless case by
$$\eqalign{&\bpa(\j) + i\pa \bar A +i[\j,\bar A] =0\cr
&\pa(\bj)  -i\pa (h\bar A h^{-1}) =0\cr}\eqn\misheom$$
where a gauge $A=0$ has been chosen.
Notice that the second equation can be derived from the first one by
multiplying it with $h$ from the left and $h^{-1}$ from the right.
A similar  result but with $ h_{ad}$ replacing $h$  follows the
variation of eqn.  \mishwzwad\   with respect to $ h_{ad}$.

As was discussed in the introduction our aim is to analyze
the system in the presence of external sources.
External  currents are coupled to the system by
adding to the action \mishwzwghl\ or \mishwzwad\  the
following term
$${\cal L}_{ext}={1\over
2\pi} \int d^2 z Tr(J_{ext} \bar A + \bar J_{ext}  A).$$
The variation of the combined action with respect to $A$ and $\bar A$
(and then setting $A=0$)  yields
 for the case of $N_f$ fundamentals
the following equations of motion
$$\eqalign{&\pa^2 \bar A  + \alpha_c(iN_f\j+J_{ext}) =0\cr
&\pa\bpa \bar A +[i\pa \bar A, \bar A] - \alpha_c[N_f(i\bj+h \bar A h^{-1}
-\bar A)+\bar J_{ext}]
=0\cr}\eqn\misheomA$$
where $\alpha_c={e_c^2\over 4\pi}$.
It follows from the equations of motion \misheom\ and
\misheomA\ that  both the dynamical currents $j_{dy}={iN_f\over 2\pi}\j,\
\bar j_{dy}={iN_f\over 2\pi}[\bj -ih \bar A h^{-1} +i\bar A]$ as well as the
external currents are covariantly conserved, which for
$A=0$ reads  $$\bar D j_{dy} +\pa \bar j_{dy}= 0\qquad
\bar D J_{ext} +\pa \bar J_{ext}= 0.\eqn\mishccc$$
with $\bar D= \bpa - i[\bar A , ]$.
One can  eliminate the dynamical current and derive the following  equation
for    the gauge fields in terms of the external currents
$$\pa\bpa \bar A +[i\pa \bar A, \bar A] + \alpha_c(N_f\bar A-\bar J_{ext})=0
\eqn\misheqA$$
In fact the equation
 one gets  in this way is the $\pa$ derivative
acting on the l.h.s  of \misheqA\  equals zero.
However, one can fix the residual gauge
invariance  $\bar A\rightarrow iu^{-1}\bpa u + u^{-1}\bar A u$, with
$u$ an anti-holomorphic function $\pa u=0$
(thus preserving $A=0$), to eliminate the antiholomorphic
function that should have been  put in  the r.h.s of \misheqA.
Thus $\bar A=ih^{-1}\bpa h$ and
$\bar j_{dy}=-{N_f\over 2\pi} \bar A =-{iN_f\over 2\pi}h^{-1}\bpa h$. Note that
this is not the current of the free case which is $ \bar j_{free}=
{iN_f\over 2\pi}h\bpa h^{-1}=-{iN_f\over 2\pi}(\bpa h) h^{-1}$
%so that  and to get
%\misheqA\  one needs a boundary condition, thus fixing the residual gauge
%invariance.

 The last equation, which
holds for the massless cases, \mishwzwghl\ and $m=0$ of
\mishwzwad, is universal in the
sense that it is independent of the representation of the dynamical
fermions\refmark\KuSc.
%To obtain the equation for level $k$
%for example, change $\alpha_c$ to $k\alpha_c$
%and $\bar J_{ext}$ to ${1\over k}\bar J_{ext}$.
 Once mass terms are added there is an explicit dependence on the
dynamical fermion and instead of \misheqA\
one finds that  the $\pa$ derivative of its  left hand side
 is  equal to $i {m^2}N_f\alpha_c(h-h^{-1})$, where $m^2$ is given in
eqn. \mishcb\ and where for the fundamental representation
we have chosen non-flavored configurations namely we
set $g,l,\phi$ of eqn.\mishwzwghl\ to $1,1,0$ respectively
(otherwise instead of a factor $N_f$, $h$ is multiplied by the flavored
contribution $Tr
gle^{i\sqrt{4 \pi\over N_CN_f}\phi}$).

Studying  the  quantum system by  analyzing the corresponding equations of
motion is a justified approximation only provided that the classical
configurations dominate the functional integral. Such a scenario can be
achieved for the case of
massless quarks in the fundamental color representation in the limit of a
large number of flavours. In fact one can show that in that case ${1\over
N_f}$ plays the role of $\hslash$. The colored sectors of the action
takes the form
$$S =N_f\{S_{1}(h) -{1\over
2\pi} \int d^2 z Tr(ih^{-1}\pa h \bar A )
+{1\over \tilde e_c^2}\int d^2 zTr[\pa \bar A]^2\} \eqn\mishlnf$$
where $\tilde e_c= e_c\sqrt {N_f}$.

\section{ Solutions of the equations without external quarks}

Let us consider first
the case where the exteral sources are switched off.
 It is obvious
that an ``abelian"  massive mode is a solution of the
equations of motion \misheqA. Consider a configuration of the form
$\bar A\equiv T^a \bar A^a(z,\bar z)= T^a\delta^{a,a_0} \bar {\cal A}(z,\bar
z)$,where $a_0$ is a given index that takes one of the values
$1,...,N_c^2-1$,  then the commutator term
vanishes and  ${\cal A}$ has to solve
$\pa\bpa \bar {\cal A} + \tilde\alpha_c\bar{\cal A}=0$
with $\tilde\alpha_c= N_f\alpha_c$.
It is clear that there are $N_c^2-1$ such solutions and in fact it is easy to
see that this property will be shared by every possible solution. This
follows from the fact that the equation of motion is not invariant but rather
covariant with respect to the ``global color" transformation
$A\rightarrow  u^{-1}A u$ with a constant $u$.

Let us now check whether the equations admit soliton solutions.
For static configurations the equation reads
$$\pa_1^2\bar A -\sqrt{2}[i\pa_1 \bar A, \bar A] -2 \tilde\alpha_c\bar A=0
\eqn\misheqAs$$
Multiplying the equation by $\bar A$, taking the trace of the result and
integrating over $dx$ one finds after a partial integration that
$\int dx[ Tr [(\pa \bar A)^2 + 2\tilde\alpha_c A^2]=0$
which can be satisfied only for a vanishing $\bar A$.

In the search for other possible solutions one may be instructed by the fact
that $F_{z\bar z}=\pa \bar A$ can be written in 2d as $F_{z\bar z}=
\epsilon_{z\bar z} F$ and impose an ansatz for the solution of the form
$\bar A = zC(\rho)$ where $\rho=z\bar z$. Expanding $C$ as a power series in
$\rho$ one finds that the commutator term has to vanish and $C$ is determined
by the equation $ \rho C" + 2C' + \tilde\alpha_c C =0$, where $C'=\pa_\rho C$.
By
a change of variable $C= \rho^{\half} W(x)$ one can rewrite this
equation as  $ W" +{1\over x} W' + (1-{1\over x^2})W=0$
with $x^2=4\tilde\alpha_c\rho$, which is $x^{-2}$ times  a Bessel
equation of order one, so that the solution for the gauge field takes the form
$$ \bar A = {\bar A_0z\over \sqrt{ \alpha_c z\bar z}} J_1(2\sqrt { \alpha_c
z\bar
z})\eqn\mishBes$$
where $\bar A_0$ is an arbitrary constant matrix.
 %The energy that is associated with this solution is finite and given
%by...

The next task is to examine whether there are any possible solutions which are
``non-abelian" in their nature.
Consider in  the special case of
$SU(2)$  the configuarion $\bar A = e^{-i\theta \tau_0}\bar A_0 e^{i\theta
\tau_0}$ with  a constant matrix  $\bar A_0=e_0\tau_0+\bar e\tau+e\bar\tau$
. Plugging this ansatz into eqn.
\misheqA\ with no external source
 one finds that there is a solution provided that $\pa \theta$ and
$\bpa\theta$ are constants, namely
 $\theta= \theta_0 +k\bar z+\bar k z$
where $k$,
  $\bar k$ and $\theta_0$ are constants. Indeed the following
gauge field

$$\bar A ={\tilde\alpha_c-k\bar k\over \bar k}\tau_0 +
\sqrt{(k\bar k-\tilde\alpha_c)\tilde\alpha_c\over 4\bar k^2}[e^{-i\theta}
\tau+e^{i\theta} \bar \tau]\eqn\mishsolu$$ is a `` non-abelian solution".
The notation and the derivation are presented in appendix A.
Setting $\theta_0=0$ requires that $(k\bar k-\tilde\alpha_c) >0$.
Looking into the case where $\bar A_0$ is not a constant, but with
$\pa\bar A_0=0$ to preserve $F(\bar A_0)=0$, one gets that the only solution is
 with a constant $\bar A_0$

%The ansatz of above with a constant $\bar A_0$ is a special case of a
%transformed (with $e^{i\theta\tau_0}$)  flat gauge connection $F=0$. In fact
%it is easy to check that from the general solution of $F=0$, namely $\bar
%A(\bar z)$, it is only ( without loss of generality)
%the    constant  $\bar A_0$ that leads to a solution.
%One can allow a $\bar z$ dependent phase in the coefficients of $\tau$ and
%$\bar \tau$, but this can be absorbed in $\theta$.
%Assuming a general $\bar z$ dependence of $\bar A_0$ leads only to the above
%solution.

The solution we have found eqn. \mishsolu\ are truly non-abelian solution.
The corresponding $F$ is $F=
 i\bar k e^{-i\theta \tau_0}(\bar e\tau-e\bar\tau)e^{i\theta \tau_0}$.
Performing a gauge transforamtion with $U=e^{-i\theta \tau_0}$ one finds that
$F_U= i\bar k(\bar e\tau-e\bar\tau)$, $\bar A_U =\bar A_0+k\tau_0$ and
$A_U=\bar k\tau_0$ (recall  that $A=0$).  We thus found that $A_U,\bar A_U$ and
$F_U$ are fixed in space-time and no two commuting. Furthermore an abelian
gauge configuration of the form $\bar A=-i\bar kz(\bar e\tau-e\bar\tau)$ and
$A=0$ that leads to the same $F$ is not connect to $A_U,\bar A_U$ by a gauge
transformation.

%One may try to generalize the ansatz of above and try a transformed
%configuration of the form
%$\bar A = e^{i\omega^a \tau_a}\bar A_0 e^{-i\omega^a
%\tau_a}$. This will lead to the same solution as eqn. \mishsolu\.???

Using the expression of $\bar A$ one can easily
extract $j_{dy}$ and $\bar
j_{dy}$. This will be done in section 6. Moreover, one can determine  the
non-abelian group factor $h$. From eqn.\misheomA\ it follows that
$h^{-1}\pa h =i{\pa^2\bar A \over \tilde\alpha_c}$. Using the ansatz for $\bar
A= e^{-i\theta \tau_0}\bar A_0 e^{i\theta
\tau_0}$ it is easy to find that
$$ h= e^{-i{\bar k z }[\tau_0 + \tilde e(\tau +\bar
\tau)]} e^{i\theta \tau_0}\eqn\mishh$$
where $\tilde e=\half\sqrt{{\bar k k \over \tilde\alpha_c}-1}$.
%A different derivation of $h$ is presented in  the appendix.

\section{ Solutions of the equations with external current}

Next we want to turn on a  covariantly conserved (eqn. \mishccc)
 external current $J_{ext}$ and study the
corresponding equations of motion.
Abelian solutions are easily constructed. For instance for a pair of quark
anti-quark as an external classical  source
$\bar J^a{ext}=T^a\delta^{a,a_0}Q[\delta(x_1-R)-\delta(x_1+R)]$
the abelian solution is
$$ \bar A=\half\sqrt{2\alpha_c}T^a\delta^{a,a_0}
Q[e^{-\sqrt{2\tilde\alpha_c}|x_1-R|}-
e^{-\sqrt{2\tilde\alpha_c}|x_1+R|}].\eqn\mishabexso$$
Inserting this expression into ${1\over 2\pi}\int dx_1Tr[A\bar J_{ext}]$ one
finds the usual screening potential\refmark\GKMS 
$$V(r)= {1\over
2\pi}\sqrt{2\alpha_c}Q^2(1-e^{-2\sqrt{2\alpha_c}|R|})Tr[(T^{a_0})^2]
\eqn\mishVR$$.

Again the challange is to find ``non-abelian" solutions where the commutator
terms do not vanish.
The  $SU(2)$ ``non-abelian solution" of above is a solution also in case of
a constant external current  $\bar J= \tau^a\delta^{a,0}J_0$ with the trivial
modification that
$g$ ( see appendix A) is replaced by  $\sqrt{\tilde\alpha_c [{J_0\bar k\over
N_f}+(k\bar k-\tilde\alpha_c)]\over 4\bar
k^2}$.
%where $ge^{-i\theta}$ is the coefficient of $\tau$ in $\bar A$ (see also
%appendix).
 Consider now an external current of the form
$\bar J= \bar J_0(\bar z) \tau_0 $.
A solution in that case is
$$\bar A=(f_0+J_0(\bar z))\tau_0 +
[g(e^{[-i(k\bar z + \bar
k z+I)]})\bar \tau+c.c]$$
where $\bpa I(\bar z)={1\over N_f}\bar J_0(\bar z)$ with
 $f_0$ and $g$  related to $k$ and $\bar k$ as given in eqn.
\mishsolu.  In the case of light-front quatization with $\bar z$
playing the role of the space coordinates, $\bar J_{ext}(\bar z)$ stands for a
general ``static" current. In particular a current density that corresponds
to a quark anti-quark pair takes in this framework the form
$\bar J^a_{ext}=\half\tau^a\delta^{a,a_0}Q[\delta(\bar z-R)-\delta(\bar z+R)]$
and the corresponding solution has $\epsilon(\bar z-R)$
and  $\epsilon(\bar z+R)$ factors in $\theta$.
The corresponding potential is a constant thus  non-cofining.

  \section{The energy-momentum tensor and the spectrum}
Next we want to analyze the spectrum of physical states
that
 correspond  (at least in the large $N_f$ limit)
to solutions of the equations of motion.
Recall that  those states
 transform in the adjoint representation of the global color
transformations.
%The picture that associates with the abelian solutions  is clear and thus
%we proceed to analyze the non-abelian solution of eqn. \mishsolu.

First  we have   to compute the  energy momentum tensor $T\equiv
T_{zz}, \bar T\equiv T_{\bar z\bar z}, T_{\bar z z}$  that corresponds to
the action \mishwzwghl.
Only the colored part of the energy momentum tensor is relevant to our
discussion. From appendix B
%The components of the latter  take the form
$$T={\pi\over N_f+N_c}:Tr[j_{dy}j_{dy}]:\ \
\ \ T_{\bar z z}= {1\over 8\pi\alpha_c}
 Tr[(\pa\bar A)^2]
\eqn\mishemt$$ where the currents of the dynamical quarks which
were defined below eqn.\misheomA\ are
$$\eqalign{j_{dy}=&{iN_f\over 2\pi}h^{-1}\pa h= -{1\over 2\pi\alpha_c}\pa^2\bar A;\ \ \
\cr
  \bar j_{dy}=&{1\over 2\pi\alpha_c}(\pa\bpa\bar A +i[\pa  \bar A,\bar A])=
  -{iN_f\over 2\pi}h^{-1}\bpa h=-{N_f\over 2\pi}\bar A \cr}\eqn\msihJdy$$
%$T$ was obtained by replacing derivatives with covariant
%ones, as no contribution from the pure gauge part. Similar derivation
%applies for $\bar T$. $T_{\bar z z}$ comes from the gauge part only.

To procced and compute the masses of the physical
states one has to choose a quantization
scheme. It is natural in the light cone gauge to use a light front
quantization.  In that scheme we take $ z$ to denote the space
coordinate. In appendix B we express the
  momentum coponenets
$\bar P$ and $ P$ as integrals over $T$ and $T_{\bar z z}$.
 The  masses of the states are given by the eigenvalues of
$M^2=P\bar P$.

To set the proper normalization of the fields let us consider
first the abelian solutions for $\bar A$.  In that case the operator
$\bar A$ can be written as

$$ \bar A_{ab} = T^Ie_c\int {d k\over {\cal N}(k)}[ a(k) e^{-i(k\bar z +\bar k
z)} +c.c]$$
where $T^I$ is a matrix in the Cartan sub-algebra,  $k\bar k=\tilde\alpha_c$,
the creation and annihilation operators obey the commutation relation
$[a(k),a^\dagger(\tilde k)]=\delta(k-\tilde k)$ and ${\cal N}(k)$
is a normalization factor. Inserting this form of $\bar A$ it is a
straightforward calculation to get ${\cal N}(k)= 2\sqrt{\pi k}$
so that $M^2$ on the states $|k,\bar k>$ is equal as expected to
$\tilde\alpha_c$.

One can  instead assume a finite system of size $L$ in $ z$
direction. In that case if one    uses
a normalization where $\bar k L$
is an integer $n$ times $2\pi$, and that   $\bar A_{ab} ={2\over \sqrt{n}} k
sin \theta$ (in first quantized version).
One then finds  
%in the large $ L$ limit
  that $ P=  k$ and $\bar P= 
 \bar k$, so that  again $M^2=\tilde\alpha_c$.

%One can also workout the expression for $M^2$ for the Bessel funcion solution
%of eqn. \mishBes\...

In case of the $SU(N_c=2)$ non-abelian solutions the superposition
principle  does not apply  and  there is no room for Fourier expansion of
fields. We thus invoke the second quantization method of above, with
  a finite size system.
The expectation values of the energy momentum tensor in those states
are determined by
substituting the  expression for the  gauge
configuration $\bar A$ into eqn.\mishemt.
This leads  to
$$\eqalign{<T>&={1\over 8\pi}{N_f^2\over N_f+N_c} (\bar k)^2
 { k\bar k\over \tilde\alpha_c}(1-{\tilde\alpha_c\over k\bar k})\cr
<T_{\bar z z}>&={N_f\over 16\pi} (k\bar k) 
 (1-{\tilde\alpha_c\over k\bar k})\cr}
\eqn\mishemt$$
If we use  the same quantization scheme as for the abelian case we
get that
$$ \bar P={1\over 8\pi}{N_f^2\over N_f+N_c}( L \bar k) \bar k
 { k\bar k\over \tilde\alpha_c}(1-{\tilde\alpha_c\over k\bar k})\ \ \
\  P={N_f\over 16\pi}( L \bar k)  k
 (1-{\tilde\alpha_c\over k\bar k})$$
where $ L$ is the size of the system. Taking  again the
 normalization  $\bar k L = 2\pi n$
 one finds that the non-abelian state is
characterized by  masses
$$M^2={n^2\over 32}{N_f^3\over N_f+N_c}
 (k\bar k){ k\bar k\over \tilde\alpha_c}(1-{\tilde\alpha_c\over k\bar k})^2$$
The discussion  above was all for $\theta_0=0$.
Thus  we require that $(k\bar k-\tilde\alpha_c)>0$\ \ \  
(the case of zero $k\bar k-\tilde\alpha_c$ corresponds to vanishing $\bar A$).
We get an $M$  starting from $M=0$  and
growing up linearly in $k\bar k$ for $k\bar k>>\tilde\alpha_c$.
Note also that our solution is singular for $\tilde \alpha_c=0$.

\section{ Bosonized external currents}

Another approach to the coupling of the dynamical fermions to external
currents is to bosonize the ``external" currents.
Let us briefly summarize first the abelian case. Consider external
fermions of mass M and charge $qe$ described by the real scalar filed $\Phi$
together with the dynamical fermions of unit charge $e$ and mass $m$ associated
with the scalar $\phi$.  The Lagrangian of the combined system after
integrating  out the gauge fields is given by
$$\eqalign{{\cal L}=& \half(\pa_\mu\phi\pa^\mu\phi) +
m\Sigma[cos(2\sqrt{\pi}\phi)-1]
+ \half(\pa_\mu\Phi\pa^\mu\Phi) +
M\Sigma[cos(2\sqrt{\pi}\Phi)-1]\cr
&-{e^2\over 2\pi}(\phi + q\Phi)^2\cr}\eqn\mishbosex$$
Let us look for
static solutions of the corresponding equations of motion with
 finite energy. Take, without loss of generality,
 $\phi(-\infty)=\Phi(-\infty)=0$. From the M term we get
$\Phi(\infty)=\sqrt{\pi}N$ with $N$ integer.  For $m\ne 0$ we also get
$\phi(\infty)=\sqrt{\pi}n$. Now from the $e^2$ term,  $n+qN=0$.
Thus, for instance for    $N=1$ finite energy solutions occur only for
$q=-n$. In the massless case we have only $\Phi(\infty)=\sqrt{\pi}N$
and then  a finite energy solution for $N=1$ is if
 $\phi(\infty)=-\sqrt{\pi}q$.
So, when $q=-n$, the system is  always in the screening phase, whereas when
$q\ne -n$ it is in the confinement phase  for $m\ne 0$ and screening for
$m=0$.

Proceeding now to the QCD case  one can
consider  several different possibilities $(J_{ext}^{F},j_{dy}^{F}),
(J_{ext}^{F},j_{dy}^{ad}),(J_{ext}^{ad},j_{dy}^{F}),(J_{ext}^{ad},j_{dy}^{ad})$
and with dynamical fermions that can be either massless or massive.
Obviously, for the external source one would assign a mass which should then be
taken to infinity.
The system of dynamical adjoint fermions and external fundamental quarks can
be described by an action which is the sum of  \mishwzwad\ and \mishwzw.

Integrating over the gauge degrees of freedom one is left with the
terms in \mishwzw\ and  \mishwzwad\ that do not include coupling to gauge field
together with a current-current non-local interaction term.
For the interesting case of dynamical quarks in the adjoint and external in
fundamental we get for the interaction term
$$\int d^2 z \sum_a\{{1\over \pa}[Tr(T^a_Fi u_{ext}^{-1}\pa u_{ext})+
{i\over 2} Tr(T^a_{ad}\j )]\}^2\eqn\mishscr$$
where $u$ is defined in \mishwzw,
$T^a_F$ are the  $SU(\nc)$ generators expressed
as $(\nc\nf)\times(\nc\nf)$ matrices in the fundamental representation
of $U(\nc\times \nf)$
and $T^a_{ad}$ the
$(\nc^2-1)\times (\nc^2-1)$  matrices
in the $SU(\nc)$ adjoint representation. The other cases can be
treated similarly. For simplicity we discuss from here on the case of a single
flavor.

Let us first  consider the
case of external adjoint  
quarks $u_{ext}(x)\in SU(\nc)\times U_B(1)$, which can be
represented as 
 $$u_{ext}\,=\,\pmatrix{\matrix{e^{-i\Phi}&&&&\cr
&e^{i\Phi}&&&\cr &&1&&\cr &&&\ddots&\cr&&&&1\cr}}\eqn\mishstso $$
($\Phi$ is not normalized canonically here). 
The reason that we take a diagonal ansatz is that it corresponds,
as we argue in  appendix C, to
a minimal energy configuration.
Ansatz \mishstso\ corresponds to $Q^1_{ext}\bar Q^2_{ext}$, namely,
to an external adjoint state. We expect this state to be screened
by the adjoint dynamical fermions. With this ansatz  ${1\over
\pa}(ih_{ext}^{-1}\pa h_{ext})$ takes the form
$$\pmatrix{\matrix{\Phi&&&&\cr &-\Phi&&&\cr &&0&&\cr
&&&\ddots&\cr&&&&0\cr}}\eqn\mishpastso $$ It is thus clear that
only $T^3_F$ contributes to the trace in \mishscr. To show the
dynamical configuration that screens, take 
$$log\ 
h_{ad}=\pmatrix{\matrix{0&-\phi&0&&\cr \phi&\  0&0&&\cr
0&0&0&&&\cr &&&&\ddots&\cr&&&&&0\cr}}\eqn\mishpastsoh $$ with
the matrix that contributes  to the $Tr$ in \mishscr,
$$ T_{adj} = i\pmatrix{\matrix{\ \ 0& 1& 0&&&\cr
-1& 0& 0 &&&\cr\ \  0& 0& 0&&&\cr&&&0&&\cr
&&&&\ddots&\cr&&&&&0\cr}}\eqn\mishpastso $$ 
which corresponds to the generator of rotation in direction 3
for the sub $O(3)$ of first three indices, 
thus obtaining the a term
proportional to  $(\Phi+\phi)^2$  emerging from \mishscr. The mass terms for
$u_{ext}$ and $h$ are now proportional to  $(1-cos\Phi)$ and 
$(1-cos\phi)$ respectively.  A boundary condition
$\Phi(\infty)=2\pi$ can be cancelled in the interaction term
 by the boundary condition $\phi(\infty)=-2\pi$.

Let us examine now the case of a single external quark
$$u_{ext}\,=\,\pmatrix{\matrix{e^{-i\Phi}&&&\cr
 &1&&\cr &&\ddots&\cr&&&1\cr}}\eqn\mishstso $$
Its  contribution  to  the  interaction term
is $\sum_{i=1}^{\nc-1}(\eta_i\Phi+``dyn")^2$ where
%\sum_{i=2}^{\nc}
$\eta_i=
{1\over
\sqrt{2i(i+1)}}$ and $``dyn"$ is the part of the dynamical quarks.
If again we take a configuration of the dynamical quarks based
on a single scalar like in \mishpastsoh\
we get altogether an $e^2$ term of the form
$(\half\Phi+\phi)^2$.  
% where $\sigma=...$.
%In the simpest case of $\nc=2$ one finds $\eta=\half$ and
%$\sigma=1$. 
Now if $\Phi(\infty)=2\pi$ one cannot find a finite
energy solution since from the mass term  $\phi(\infty)=2\pi n$,
and thus there is no way to cancell the interaction term. If,
however, we consider massless dynamical fermions there is no
constraint on   $\phi(\infty)$ so it can be taken to be equal
$-\pi$, and thus  again a screening situation is achieved.
This argument should be supplemented  by showing that one cannot
find another configuration besides \mishpastsoh\  that may cancel
 the $\eta_i\Phi$ term in $(\eta_i\Phi+``dyn")^2$, for
$SU(N_c)$ with $N_c\geq 3$. 
\section{Discussion}

In the present paper, using bosonization techniques, we have
presented further evidence for the non-abelian
screening of external charges by
dynamical massless fermions. We have shown it explicitly for 
dynamical fermions
in the adjoint and  $N_f$ fundamental representations. In fact, the
latter case implies that a WZW model coupled to non-abelian gauge fields
is a screening model for any level of the affine Lie algebra.
The fact that there is no relation between the charges of
the screening dynamical fermions and those of the external sources may seem
unintuitive. However, one can understand this phenomena in a simple
way if one realizes that the interaction between the external
charges involves an exchange of a massive mode
which is an outcome of the dynamics of  bare massless quarks.
The main outcome  of  the present paper is the observation that
indeed such massive modes manifest themselves in the form of
soutions of the equations of motion.
This is well known for $QED_2$, and has been emphasized more recently for
$QCD_2$.\refmark\FHS

  A natural question to ask
is whether there are consequences 
   of the screening behaviour in the spectrum
of the theory. A simple minded intuition  of the difference between
a confining and a screening spectrum can be derived from quantum
mechanics. A potential of the form \mishVR\ leads to a spectrum of
bound states with energy smaller than  the
asymptotic value of the potential. A linear potential, on the other hand,
can accomodate an infinite spectrum of bound states with no
limitations on their energies.
Practically, of course, higher energy states will be unstable.  
It is obvious that the massless Schwinger model which has a single
state in its spectrum falls into the former class. It looks plausible
that the spectrum in the non-abelian case is also limited.

Another way to distinguish between confining and non-confining
spectrum is the dependence of multiplicity of the physical states
 on the number of colors. In a confining
spectrum one  finds only color singlets and their multiplicities do not
depend directly
on $N_c$. The massive states discussed in the present work,
both the abelian and the non-abelian solutions, admit  a degeneracy
of  $N_c^2-1$, or stated differently those states are in the adjoint of
the  ``global color symmetry".

We have used an argument that in the large $N_f$ limit the
classical  solutions
of the equations of motion dominate the functional integral. However,
it is not obvious that the corresponding states are physical.
In a previous paper\refmark\FHS
we have used a special
 formulation of \Q
 in terms of $A=if^{-1}\pa f$, $\bar A=i\bar f\bpa\bar f^{-1}$
with $f(z,\bar z)$, $\bar f(z,\bar z)\in[SU(N_c)]^c$  the complexification
of \snc . Implementing BRST techniques we have solved for the physical states
of the abelian analog.  In the non-abelian case one finds
 massive color singlet states
with a mass  of
$e_c\sqrt {N_f \over 2\pi}$ which are the analogs of the states
discussed in the present paper. Unlike the massless partners of these
states, for the massive ones we were not able to show that they are
not physical states.
We still
do not have a definite answer to this question;
however, the fact that in the abelian theory of the large $N_f$ limit
they are physical states  supports the conjecture that they are physical
also for
finte $N_f$.
Furthermore,  it seems that
no matter whether they are in the sub-space of physical states or
not they are responsible for the screening potential.

One may suspect that the massive ``Schwinger like" states
are an artifact
of the abelianization of the theory in the large $N_f$ limit.
To exclude this possibility we have found
truly  non-abelian solutions
of the equations of motions.

The fact  that an abelian nature is not  necessary for the
existense of the massive modes tells us that it is plausible that the
spectrum is characterized by
 a smooth transition of the screening behaviour
from large $N_f$  down to $N_f=1$.
 It may seem that there is a contradition between the
seminal work of 't Hooft and the various types of evidence
of the screening behavior of the massless theory. The reason
for that  is that
 the analysis of ref. [\tHooft],  which predicts a confining spectrum,
 is
insensitive to the question of whether the quarks are massless or
massive. The way to reconsile the two pictures of the massless model
is the following.  Assume that the potential is of the form of eq.
\mishVR.  In the approach of [\tHooft], $e_c^2N_c$ is kept finite in
the large $N_c$ limit. This implies that the potential behaves like
$(1-e^{-{\kappa\over\sqrt{N_c}}|R|})\sim {\kappa\over\sqrt{N_c}}|R|[(1-\half
{\kappa\over\sqrt{N_c}}|R|+o({1\over N_c})]$
for fixed $R$ and large $N_c$, with $\kappa=2\sqrt{2\alpha_c N_c}$
 a finite constant.
Now it is clear that in the limit of
$N_c\rightarrow \infty$ the potential looks like a linear potential
which obviously admits a confinement behaviour.
Thus, the large $N_c$ limit prevents one from
detecting the truely screening nature of the massless system.

In the present paper we have not discussed the solutions of the
equations of motion and the corresponding potentials for the case of
massive quarks (for large $N_f$ it is discussed in [\Adi]).
 However, using the analogy with the massive Schwinger
model one can get a general picture of the passage to a confinig
behavior. The mass of the massive state of the Schwinger model is
shifted once quark mass is turned on. But an additional massless state
emerges. Exchange of the latter mode causes confinement.
Presumably a similar situation occours in the non-abelian case.
 Moreover,  in the double bosonization description
of the massive model
where   the external sources are also bosonized, the confining
nature  of the theory manifest itself via the absense of quark
soliton solutions. This fits well  with the confining spectrum
discovered by 't Hooft, the baryonic spectrum analayzed in ref. [\DFS]
and the results of ref. [\GKMS]. One can envisage having a term $m^2|R|$
in the potential, where $m$ is the quark mass, in addition to the screening
term, with only a screening remaining for $m=0$.  

Several open questions arise following the results of the present work, for
instance the quantization of the non-abelian solutions of the equations of
motion, deriving  similar solutions for the massive model, etc.
In the introduction the long standing puzzle of confinement in non-abelian
gauge theories in 4D was mentioned as one of the motivations for the present
work, so we cannot end this discussion without addressing the question of
the relevance of our results to the real world QCD.
The screening of  external quarks  in the fundamental representation by
dynamical fermions in the same representaion both in the massless and massive
cases seems to fit the picture one has about 4D QCD with fundamental quarks.
The analysis of 2D model with adjoint fermions exhibit a screening behaviour
for massless quarks and confinement one for massive case. In 4D one believes
that adjoint quarks cannot screen external fundamental ones.  The mass of the
dynamical fermions does not  play  any role in this issue 
in 4D. This phenomenon that the
 nature of the mass term in 2D is very different than  the one in 4D, was
found also in other circumstances like the baryonic spectrum.\refmark\DFS
One may speculate that the analog of quarks in 4D, whether massive or
massless, are massive quarks in 2D. In [\GKMS] it is speculated that the
analog of the phase transition that occours by turning on mass in the adjoint
case is that of breaking SUSY and loosing screening in $N=1$ supersymmetric
YM in 4D.

\ack{ One of us (J.S) would like to thank A. Zamolodchikov for numerous
fruitful discussions}.

\refout

\Appendix{A}
\centerline{Non-abealian
solution of the equation of motion for $SU(2)$}

Let us parametrize the SU(2) gauge field in the following form
$\bar A=f\bar\tau + \bar f\tau +f_0\tau_0$ where $f_0$ is real, $\bar f$ is the
complex conjugate of $f$ and    the matrices $\tau$ obey the
algebra $[\tau_0,\tau]=\tau;\ [\tau_0,\bar\tau]=-\bar\tau;\
[\tau,\bar\tau]=2\tau_0$. In terms of those variable the equation of motion
eqn (8)  read ( with $\bar J_{ext}=0)$
$$\eqalign{ (\pa\bar\pa +\tilde\alpha_c)f &- i(f\pa f_0-f_0\pa f)=0\cr
(\pa\bar\pa +\tilde\alpha_c)f_0 &+ 2i(\pa \bar f f-\bar f\pa
f)=0\cr}\eqn\mishsu $$
Using the notation $f=ge^{i\theta}$ eqn. \mishsu\ takes the form

$$\eqalign{ (\pa\bar\pa +\tilde\alpha_c)g & -g\pa\theta(f_0+\bpa\theta) =0\cr
g\bpa\pa\theta +\bpa \theta\pa g&+ \pa \theta\bpa g-(g\pa f_0 -f_0\pa g)=0\cr
(\pa\bar\pa +\tilde\alpha_c)f_0 &+ 4\pa\theta g^2=0\cr}\eqn\mishsuu $$
%Let us first consider solutions with $g$ and $f_0$ constants.
 The case where
  $\bpa\theta=\pa\theta=0$ is the abelian solution since all the
commutator terms vanish.
In case that
  $\bpa\theta=k ;\pa\theta=\bar k$
  where $k$ and $\bar k$ are constants it is easy to check that the equation
do admit a non-abelian solution  as eqn. (12)  of the form
$$\theta= k\bar z+ \bar k z;\ \ f_0={\tilde\alpha_c-k\bar k\over \bar k};\ \
g^2={(k\bar k-\tilde\alpha_c)\tilde\alpha_c\over 4\bar k^2}.\eqn\mishsol$$
This result follows also from the ansatz
$\bar A = e^{-i\theta \tau_0}\bar A_0 e^{i\theta
\tau_0}$ ($\bar A_0$ constant) for which eqn. (8)  takes the form
$$-\pa\theta\bpa\theta[\tau_0,[\tau_0,\bar A_0]] +i\pa\bpa \theta [\tau_0,\bar
A_0] + \tilde\alpha_c \bar A_0 +\pa \theta [[\tau_0,\bar A_0],\bar A_0]
=0\eqn\mishansa$$ inserting the values for the commutators one finds
$ \tilde\alpha_c f_0+4\bpa \theta g^2=0$, namely, $\bpa\theta=k $
where $k$ is a constant. Using this
property leads to a similar relation  $\pa \theta=\bar k$ where
 $\bar k$ is  also a constant and to the determination of $g$ and $f_0$ as
given in eqn. (12).
%\mishsolu.
When an external source of the form $\bar J_{ext}=\tau_0\bar
J_0$ is added the only modification of
eqn. \mishsuu\ is that  the r.h.s of the last equation 
becomes $\alpha_c\bar J_0$. Similarly, the r.h.s. of (A.4) becomes
$\alpha_c\bar J_{ext}$. When $\bar J_0$ is not a constant, but a
function of $\bar z$, we need to add to $\theta$ a function of 
$\bar z$, as discussed in section 5. 

\Appendix{B}

\centerline{Derivation of the Energy-momentum tensor of $QED_2$}
In order to understand  the energy-momentum tensor of   $QCD_2$, we start
with the derivation in the abelian theory.
The Lagrangian density  of massless $QED_2$ is given 
in bosonized form by
$$  {\cal L}_{QED}=\pa\phi\bpa\phi + \half F^2 + {e\over \sqrt{\pi}}
[\pa \phi \bar A -\bpa\phi A]\eqn\mishaction$$
where $F=F_{z,\bar z}=\pa\bar A-\bpa A$.
The corresponding equations of motion are
$$2\pa\bpa \phi =-{e\over \sqrt{\pi}}F;\ \ \ \pa F = {e\over \sqrt{\pi}}\pa\phi;
\ \ \ \bpa F = {e\over \sqrt{\pi}}\bpa\phi \eqn\misheQ$$
We have here the currents $j_{[U(1)]}=-{1\over \sqrt{\pi}}\pa\phi$
and 
 $\bar j_{[U(1)]}={1\over \sqrt{\pi}}\bpa\phi$.
The definition of the energy momentum is
$$\eqalign{ T_{\bar z z } &= {\pa {\cal L}\over \pa(\pa \phi)}(\pa \phi)+
 {\pa {\cal L}\over \pa(\pa \bar A)} (\pa \bar A)- {\cal L}\cr
&= \pa\phi\bpa\phi + F(\pa \bar A) + {e\over \sqrt{\pi}}
\pa \phi \bar A -{\cal L}\cr
&=\half F^2 +\bpa (FA)\cr}\eqn\mishTT$$
In a similar way we have that
$$\eqalign{  T_{z\bar z }&=\half F^2 -\pa (F\bar A)\cr
 T_{\bar z \bar z }&=\bpa\phi\bpa\phi
+{e\over \sqrt{\pi}}
\bpa \phi \bar A\cr
 T_{ z z }&=\pa\phi\pa\phi
-{e\over \sqrt{\pi}}
\pa \phi A \cr} \eqn\mishTt$$
Note that we have here the ``canonical" energy-momentum, not
the symmetric one, but it generates too the Poincare group.   

In the gauge $A=0$ the above expressisons become
$$\eqalign{  T_{\bar z z }&=\half F^2; \ \ \ \ T_{z\bar z }=\half
F^2 -\pa (F\bar A)\cr
 T_{\bar z \bar z }&=\bpa\phi\bpa\phi
+{e\over \sqrt{\pi}}
\bpa \phi \bar A\cr
 T_{ z z }&=\pa\phi\pa\phi  \cr} \eqn\mishTt$$
We now take $z$ as our ``space" variable and $\bar z$ as ``time".
Thus
$$\eqalign{ P=& \int dz  T_{ z z } =\int  dz (\pa \phi)^2
=\pi\int  dz j_{[U(1)]}^2 \cr
\bar P=& \int dz  T_{ z \bar z } =\int  dz \half
F^2\cr}\eqn\mishPP$$  

In the non-abelian case, in the gauge $A=0$, $T_{ z z }$
has no contribution from $u^{-1}\pa u \bar A$ term nor from the
$F^2$ term. Thus it is the same as the one derived from the 
ungauged WZW action, namely
$$T_{ z z }= {\pi\over N_f+N_c}:j^2:$$
$T_{\bar z z }$ has no contribution from the WZW part of the
action, and the contribution from $u^{-1}\pa u \bar A$
is cancelled by the same term in $-{\cal L}$. Thus there is a
contribution only from $F^2$ term, ie. 
$$T_{\bar z  z }=\half F^2 $$
The corresponding light-cone component $P$ is given by
$$P= \int dz  T_{ z z } =\int  dz {\pi\over N_f+N_c}:j^2:\ .$$ 
For $\bar P$ we need $T_{ z \bar z }$ since 
$\bar P =  \int dz  T_{ z \bar z }$, but as the difference with 
the symmetric energy-momentum tensor does not contribute to $\bar
P$ we also have  
$\bar P =  \int dz  T_{  \bar z  z }= \int  dz \half
F^2$. 
Note from the explicit expression for the abelian case that indeed
$T_{ z \bar z }$ differs from $\half F^2$ by a $\pa$ term that 
does not contribute to the integral  defining $\bar P$.

\Appendix{C}

\centerline{ Minimum of energy for soliton solutions}

The expression of the Hamiltonian for static configurations of the
(ungauged) massive WZW model is
$$ H=\int d^2 z Tr[\pa_i g\pa_i g^\dagger] + m^2
Tr[2-(g+g^\dagger)]\eqn\mishH$$
This is also the massive non-linear $\sigma$ model. 
Consider a  diagonal solution $g_0$ of the equation of motion.
A general
non-diagonal configuration can be written as $g=Bg_0B^\dagger$ where
$B$ is a $U(N)$ matrix that is time independent. Substitution of $g$ into
\mishH\ one dinds
$$ H=\int d^2 z Tr[(\pa_ig_0+ [B^\dagger\pa_i B , g_0])
(\pa_ig_0^\dagger+ [B^\dagger\pa_i B , g^\dagger_0])+ m^2
Tr[2-(g_0+g_0^\dagger)]\eqn\mishHg$$
The difference between the latter Hamiltonian and that which corresponds
to $g_0$ is
$$H-H_0 = \int d^2 z Tr\{[B^\dagger\pa_i B , g_0]
[B^\dagger\pa_i B , g^\dagger_0]+ 2B^\dagger\pa_i B(g_0^\dagger
\pa_ig_0 + g_0\pa_ig_0\dagger)\}\eqn\mishHh$$
 For a diagonal solution  $g_0$ the last expression reduces to
$$H-H_0 = \int d^2 z Tr\{[B^\dagger\pa_i B , g_0]
[B^\dagger\pa_i B , g^\dagger_0]\}>0\eqn\mishHhh$$
Thus if $g$ is non-diagonal, it cannot be a classical solution, as after
diagonalization to $g_0$ it will have a lower energy.

\end